\documentclass[aps,prl,twocolumn,showpacs]{revtex4}

\usepackage{graphicx}
\usepackage{amsmath, amsthm, amssymb, mathrsfs}
\usepackage[utf8]{inputenc}
\usepackage{bm}
\usepackage{comment}
\usepackage{hyperref}
\hypersetup{
    pdftitle={Crumpling Wires and Liouville Field Theory},
    pdfsubject={},
    pdfauthor={Bruno Carneiro da Cunha},
    pdfkeywords={Boundary Liouville Theory, Integrable Models, Crumpling},
    pdfpagemode=UseNone,
    plainpages=false,
    pdfstartview=FitH,
    breaklinks=true,
    colorlinks=true,
    pdfhighlight=/N,
    bookmarksopen,
    linkcolor=blue,
    citecolor=blue,
    urlcolor=blue,
    citebordercolor=1 1 1,
    filebordercolor=1 1 1,
    linkbordercolor=1 1 1,
    menubordercolor=1 1 1,
    pagebordercolor=1 1 1,
    urlbordercolor=1 1 1,
  }
\newcommand{\be}{\begin{equation}}
\newcommand{\ee}{\end{equation}}

\newcommand{\goesto}{\rightarrow}

\begin{document}

\preprint{UFPE 01-08}

\title{Crumpled Wires and Liouville Field Theory}

\author{Bruno Carneiro da Cunha}
\email{bcunha@df.ufpe.br}
\affiliation{Departamento de Física, Universidade Federal de
  Pernambuco \\
CEP 53901-970, Recife, Pernambuco, Brazil}


\begin{abstract}
The study of the packing of a length of wire in a two dimensional
domain is done using techniques of conformal maps. The resulting
scaling properties are derived through the Coulomb gas
formalism of Conformal Field Theories. An analogy is drawn
between this system and the 2D gravitational -- or Liouville -- coupling
of non-conformally invariant matter in two dimensions. 
\end{abstract}


\maketitle


{\it \bf Introduction ---} Crumpling is a general problem in Physics
in a wide range of phenomena including materials science, metal foils
and organic matter \cite{1stbatch}, which poses interesting challenges
in both theoretical and applied perspectives. Recently some attention
has been drawn to the problem of crumpling of wires, especially in two
dimensions. Relevant applications include packing of DNA strands in
viruses, general polymer chains and confinement of elastic rods
\cite{2ndbatch}.

Ideas of scale invariance have been pervasive in critical phenomena
since its inception, and in two dimensions its natural conclusion, conformal
field theory, has been a valuable tool since \cite{Polyakov:1984yq},
especially in the context of random surfaces and integrable
models. These models also provide a concrete 
realization of two dimensional quantum gravity, in its representation
as a Liouville field theory \cite{Ginsparg:1993is}. Integrability via
Toda hierarchy has exposed an exact equivalence of 2D quantum gravity
and the problem of Laplacian growth
\cite{Mineev-Weinstein:2000}. Given the lack of intuitive
understanding of quantum gravity, it
would be desirable to find more amenable problems which possess
analogies with more realistic gravity problems, such as the coupling to
non-conformally invariant matter.

In this letter we show an equivalence between the problem of
two-dimensional packing of wires and Liouville field theory
coupled to matter in an ``almost conformal'' manner. The breaking of
conformal symmetry stems from the boundary conditions and are hence
tractable by conformal perturbation theory. We also give a simple
scaling argument that predicts a phase transition where the length of
packed wire is a maximum and the critical exponent of the energy.


{\it \bf The set up}  --- Let $D$ be the interior of the container, seen
as a region of the complex plane. For simplicity, take $D$ to be the
unit disk. The packing process begins with a diameter of the disk
being continuously deformed so as to increase its length while
keeping the angles at the extremities fixed, which is called an {\it
  injection}. At a given wire length $L$, consider $z_L(t)$, the map 
that takes the initial diameter $t\in[-1,1]$ to the configuration of
the injected wire $C$. We will want to view the function $z_L(t)$ as the
restriction of a transformation (or diffeomorphism) of the whole
domain $D$ into itself. We will argue below that the condition of
non-self-intersections of the curve $C$ can be cast in terms of the
diffemorphism of the domain $D$ by saying that the regions above and
below the curve $C$, $D_+$ and $D_-$, are respectively mapped to the
upper and lower semi-disc by analytic functions. 

Such extension of $z_L(t)$ to the whole domain $D$
preserves its manifold structure and as such the function
$z_L(t)$ can be seen as locally analytic. In the next section we will
further explore this property. Also, the restriction of such maps
to the boundary of $D$, the unit circle $\mathrm{S^1}$ in our case,
forms a group, $\mathrm{DiffS^1}$ with ubiquitous applications to
conformal field theory. Seeing $D$ as a complex manifold, the
restriction enforces that we can cover the image of $D$ with an atlas
consisting of two {\it simple} analytical maps, with an ovelap region
that contains a neighborhood of $z_L(t)$. By definition, a
{\it simple map} takes simple curves to simple curves and therefore
the restriction forbids intersections \cite{Cohn}.  The transition
functions defined in the overlap will themselves be analytic in order
to warrant non-self intersection. The construction will be explicited
in the last section. 

Physically, the importance of this restriction arises as one injects
wire into the domain, or as $L$ increases. As it happens, the
system tries to minimize the energy creating points of
self-intersection.  The transition is actually akin to the
generation of vortices in the $XY$ model, in which the angle of the
tangent to the wire $\theta(t)$ goes over a full loop before reaching the other
side of the domain $\theta(t=1)=\theta(t=-1)+2\pi$. 

Let us now consider the map $z(w)$, with $L$ undetermined. The
curvature $k$ of the curve $z(t)$ and its element of length $d\ell$
are given by 
\be
k=-\frac{i}{2}\frac{\dot{z}\ddot{\bar{z}}-\dot{\bar{z}}
  \ddot{z}}{(\dot{z}\dot{\bar{z}})^2}, \quad\text{ and }\quad
d\ell^2 = \dot{z}\dot{\bar{z}}dt^2;
\ee
where $\bar{z}(t)=z^*(t)$ is the complex conjugate of the
map. The problem of \textit{adiabatic growth} is thus stated: let
$z(w)$ be the restriction of the diffeomorphism of the unit disk $D$
to $D$ defined in a neighborhood of $w$ real. The conformation of the
wire inside the domain $z(t)$ is such that the free action
\be
{\mathscr F}=\frac{\alpha}{2}\int_{-1}^{1} dt\;
\sqrt{\dot{z}\dot{\bar{z}}} 
k^2
+\mu \int_{-1}^{1}dt\;  \sqrt{\dot{z}\dot{\bar{z}}},
\label{freeenergy}
\ee
is a minimum. The Lagrange multiplier $\mu$ implements the constraint
that the minimization has to be done while keeping the total length
fixed:
\be
\frac{\partial {\mathscr F}}{\partial \mu}=L.
\label{legendretrans}
\ee
At any new injection of wire into the system, it is then assumed that
there is enough time for the elastic energy, given by the first term
of $F$, to be minimized. In practice, this time can be of the order of
several seconds for ordinary materials. The general form
(\ref{freeenergy}) is invariant by reparametrizations of the curve
$t\rightarrow t'(t)$ and as such it will be reduced to the usual forms
by a suitable choice of the parameter. Moreover, it will yield, via a
suitable change of field variables, to the relation between the
problem of crumpling and Liouville field theory.

The free energy (\ref{freeenergy}) is easily recognizable in the proper 
length parametrization $\dot{z}=e^{i\theta}$, with $\theta$ being the
angle of the curve with a fixed direction:
\be
E=\frac{\alpha}{2} \int_0^L d\ell
\left(\frac{d\theta}{d\ell}\right)^2. \label{elasticenergy1}
\ee
One can then fix the final position of the wire by means of a Lagrange
multiplier, arriving at an 
expression for the free energy: 
\be {\mathscr F}(L)= \int_0^L d\ell
\left[ \frac{\alpha}{2} \left(\frac{d\theta}{d\ell}\right)^2+ \lambda
  \cos\theta\right].
\label{freeenergy2}
\ee
Since the constraint is global, one can treat the Lagrange multiplier
as a constant coupling. This form is unsuitable to treat the various
other constraints, but it is an adequate approximation when the free
energy is dominated by the elastic bending.  The value of $\lambda$ is
determined in terms of $L$ by imposing that the endpoints are
separated by a diameter $\Delta t =2$. 


{\it \bf The Schwarz function and Liouville Theory ---} Another
interesting parametrization of (\ref{freeenergy}) is $t=z$, valid on
the curve $C$. In this case, the relevant map is $\bar{z}=S(z)$, where
$S(z)$  is called the {\it Schwarz Function} \cite{Davis}. The tangent
vector to the curve $z(t)$ is given by the relation:
\be
d\ell^2=dzd\bar{z}=S'(z)dz^2\quad \Longrightarrow\quad
\frac{dz}{d\ell}=\frac{1}{\sqrt{S'(z)}}
\ee
whereas the curvature is given by
\be
k=\frac{d\theta}{d\ell}=\frac{i}{2S'}\frac{dS'}{d\ell}
=\frac{i}{2S'}\frac{dz}{d\ell}\frac{dS'}{dz}=
\frac{i}{2}\frac{S''}{(S')^{3/2}}
\ee
One can then try to minimize the functional
(\ref{freeenergy}), but it is easier to consider (\ref{freeenergy2}),
where the constraint of constant length is already
implemented. From it results the relation $
\frac{d^2\theta}{d\ell^2}+\frac{\lambda}{2i\alpha}(
e^{i\theta}-e^{-i\theta})=0 $, which in terms of the Schwarz function is
\be
\frac{1}{S'}\{S;z\}+
\frac{\lambda}{\alpha}\sqrt{S'}-
\frac{\lambda}{\alpha}\frac{1}{\sqrt{S'}}=0,
\ee
where $\{S;z\}=S'''(z)/S'(z)-3/2(S''(z)/S'(z))^2=T$ is the {\it
  Schwarzian derivative} of $S(z)$. Upon the substitution
$S'(z)=e^{2\gamma\phi(z)}$, the equation can be written as
\be
T+V=-\frac{1}{2}(\partial\phi)^2+\frac{1}{2\gamma}\partial^2\phi-
\frac{\lambda}{4\gamma^2}e^{\gamma\phi}\left[1-
e^{2\gamma\phi}\right]=0 
\label{stress}
\ee
which is recognized as the (chiral) ``gravitational Ward identity'' of
Liouville Field theory \cite{Polyakov:1987} coupled to ``matter
field'' whose stress energy tensor is given by the exponential
terms. The constant $\gamma$ will be fixed 
using scaling arguments below. 

Strictly speaking, (\ref{stress}) is valid only on the curve $C$, but
the requisite of non-self intersection of the curve means that the
equation above should really be thought of as a analytical
continuation of the function $\phi(t)$ to a neighborhood of $C$. By
the discussion of the preceeding section, this in turn allows us to 
continue the function, and $S(z)$, into either chart of the domain
$D$. In fact, the Schwarz function does allow for a decomposition
$S(z)=S_+(z)+S_-(z)$ such that $S_\pm(z)$ is analytic at either side
of $C$ \cite{Davis}. One can further enforce that $S_\pm(z)$ thus
constructed keep the circle invariant, but for now it suffices to say that the
condition of simplicity is equivalent to the non-vanishing of $S'_\pm$
and hence that $\phi$ is well-defined.


We will be interested in the scaling properties of observables like
the elastic energy (\ref{elasticenergy1}):
\be
E=-\frac{\alpha}{4}\int_C dz \sqrt{S'}\frac{(S'')^2}{(S')^3}=
-\alpha \gamma^2\int_C dz\, 
(\partial\phi)^2e^{-\gamma\phi},\label{energy}
\ee
and the total length
\be
L=\int_C dz \sqrt{S'}=\int_C dz\, e^{\gamma\phi}\label{length}.
\ee
We note that since $C$ belongs to both charts, one can then deform the
contour to ``half a perimeter'' of $\mathrm{S^1}$, any line connecting
$w=-1$ to $w=1$, without changing the value of either observable.

Each of those observables has a definite conformal dimension, which in
this case is offset by the ``improvement term''
$\frac{1}{2\gamma}\partial^2\phi$ in (\ref{stress}). This is best dealt
in the so-called Coulomb gas formalism for Conformal Field Theories
(CFTs) \cite{DiFrancesco}.  Given the
similarity between $T$ in (\ref{stress}) and the free boson, 
let us start without the improvement term or the potential. For a free
boson  the stress energy is $T_0=-\frac{1}{2}(\partial \phi)^2$, and
the Operator Product Expansion (OPE) of correlation functions are computed
from the Green's function $\langle \phi(z)\phi(z')
\rangle=-\log(z-z')$. One can then compute the OPEs between $T$ and the
type of observables relevant to us $V_\beta(z)=e^{\beta \gamma\phi}$:
\be
T_0(z)V_\beta(z')=-\frac{\beta^2\gamma^2}{2(z-z')^2}V_\beta(z')+
\frac{\partial  V_\beta(z')}{z-z'} + \text{reg.}
\ee
where ``reg.'' stands for a regular part which implement the effect of
boundary conditions. Since $T_0(z)$ is the generator of conformal
transformations, the constants in the leading order term gives us the
behavior of $V_\beta(z')$ under scale transformations, being in fact
the operator's scaling dimensions. We then have
$\Delta[V_\beta]=-\beta^2\gamma^2/2$. Another  type of observable
important for the discussion is a {\it descendant}, which comes about
by the application of the Fourier components of $T_0(z)$ to $V_\beta$:
\be
L_nV_\beta(w)\equiv\frac{1}{2\pi i}\oint_w
\frac{dz}{(z-w)^{n-1}}T(z)V_\beta(w).
\ee
Because of the Virasoro algebra, such fields have dimension
$\Delta[V_\beta]+1$. We will be particularly interested in
$L_1V_{\beta}\equiv \beta\gamma(\partial\phi) e^{\beta\gamma\phi}$.

The usual interpretation \cite{Ginsparg:1993is} of the improvement
term in (\ref{stress}) is that the otherwise free conformal field is really
coupled to the  curvature of the base space (the $w$ plane in our
case), via the term $\phi \sqrt{g}R$. The system is no longer
invariant over fixed translations of $\phi$, but the variation is
proportional to the Gauss-Bonnet term $\int\sqrt{g}R=4\pi\chi$ and
hence a constant for fixed 
topology. This extra term will contribute to the stress energy tensor
since the latter is just the variation of the action with respect to
the metric, and $\sqrt{g}R$ does have a non-trivial variation
$\delta(\sqrt{g}R)=\sqrt{g}(g^{ac}g^{bd}-g^{ab}g^{cd})
\nabla_a\nabla_b\delta  g_{cd}$. Integrating by parts yields a
contribution by a term proportional to the second derivative of 
$\phi$, $Q\partial^2\phi$, just like the one appearing in
(\ref{stress}), where $Q=1/2\gamma$. 

Owing to the extra term in (\ref{stress}), the OPEs between $T(z)$
and the local operators $V_\beta=e^{\beta \gamma\phi}$ will be shifted, and
their new conformal dimensions will be
\be
\Delta[V_\beta]=Q\beta \gamma-\frac{\beta^2\gamma^2}{2}
=\frac{\beta}{2}(1-\beta\gamma^2).
\ee
As far as the descendants are concerned the effect of the improvement 
term is in general non-trivial, but a very fortunate coincidence
emerges for the case we are interested in: the first Fourier
coefficient of $TV_\beta$ is not affected by its presence. One can
verify this by expanding the ``kinetic'', or ``elastic'' part of the
stress energy in terms of Fourier components of the field
$\phi$. In short, the relation between the scaling dimensions of
the field and its first descendant remains valid even when the
improvement term is introduced.

We are now in position to fix $\gamma$ by requiring that both potential
terms in (\ref{stress}) have the same scaling dimensions, and substituting
values one has
\be
\Delta[V_1]=\frac{1-\gamma^2}{2},\quad
\Delta[V_3]=\frac{3-9\gamma^2}{2}, 
\ee
equating these gives us $\gamma=\pm 1/2$ and
$\Delta[V_1]=\Delta[V_3]=3/8$.  The scaling properties of the 
relevant observables $E$ (\ref{energy}) and $L$ (\ref{length}) are
dealt with by different means. $L$ is the integral of a operator of
dimension $\Delta[V_1]=3/8$, but since it is integrated over a path,
this number will be reduced by one as far as scaling transformations
are concerned: $\Delta[L]=-5/8$.
Now the integrand in (\ref{energy}) is the square of the first
descedant of the operator $V_{-1/2}$. Its dimension is then given by
$ \Delta[E]=2(\Delta[V_{-1/2}]+1)-1=7/16$,
which means that these two operators have well defined scaling
relations with respect to a change of the parameters, like $\lambda$
in the Ward identity (\ref{stress}). As there is only one dimensionful
parameter on the theory, the constant $\alpha$, 
one can summarize both behaviors in the single equation, stating how
the energy scales with $L$: 
\be
E\propto |L_{\rm crit}-L|^{-\frac{7}{10}},
\ee
where $L_{\rm crit}$ is a function of $\alpha$ and the particular form
of the domain $D$. This hints at a phase transition in which there is
a maximum length of wire that can be injected in the domain. One also
notes that given the scaling properties of $L_{\rm crit}$, it should increase
with the size of the domain, and it follows that the energy will
become more and more concentrated as the 
size of the domain increases. This is a feature the system shares with
gravity. One should note that the exponent above should be valid as
long as the system is in the elastic regime. 

{\it \bf The Boundary conditions ---} The considerations made thus far are
valid regardless of the boundary conditions. Although the details of
the constraint set the particular form of the potential in
(\ref{stress}), this is not particularly relevant for the scaling
properties studied in the last section. Here we will present a
procedure of how to deal with it from the analytical map standpoint.

Let us start from the consideration of the map $z(w)$. Since it takes
the unit disk to itself, the map $\psi(w)=-i\log z(w)$ is real on the
unit circle. It is related to the function $\phi(w)=-i\log\partial S(w)$ by
$\phi(w)=\psi(w)-i\log\partial\psi(w)+\frac{\pi}{2}$.
One would then like to impose on $\phi(w)$ constraints that reflect 
$\bar{\psi}(\bar{w}=1/w)=\psi(w)$.

The constraints are conveniently written in terms of the harmonic functions
$u=(\phi(w)+\bar{\phi}(\bar{w}))/2$ and
$v=(\phi(w)-\bar{\phi}(\bar{w}))/2i$, conjugate to each other. On the
unit circle, we then
have
\be
\begin{gathered}
u(w,\bar{w}=1/w)=\psi(w)+i\log w \\
v(w,\bar{w}=1/w)=-\log(iw\partial\psi(w)).
\label{boundarycond}
\end{gathered}
\ee
The condition that $\psi(w)$, and then $z(w)$, is locally analytic in the disk
is that $\partial\psi(w)\neq 0$ on either chart. This is tantamount to saying
that $v(w,\bar{w}=1/w)$ does not have singularities on either
chart. Combining both equations above one has $e^{-v}=-\partial_t
u+1$, where $\partial_t$ is the tangential derivative to the boundary. In
turn, $\partial_tu$ is equal to minus the normal derivative of $v$,
since they are harmonically conjugate.  It is also clear that the
numerical value $1$ represents the curvature of the boundary. Hence
the constraints satisfied by the function $v(w,\bar{w})$ can then be
accounted for by the action: 
\be
{\mathscr S}=\frac{1}{2\pi}\int_D d^2z\,\partial
v\bar{\partial}v+\frac{1}{4\pi}\oint_{\partial D}d\ell\,(vK+e^{-v});
\label{actionforv}
\ee 
where $K$ is the curvature of the boundary, constant in our case. The
area element is $d^2z=\frac{i}{2}dz\wedge d\bar{z}$. The Neumann
boundary condition that ensues from the variation of
(\ref{actionforv}) is exactly the one that arises in the
correspondence with the sum over triangulated surfaces, a discrete
version of two-dimensional quantum gravity \cite{Kostov:2003uh}.
 
The minimization of the elastic energy at constant wire length is
given by the addition of the free energy (\ref{freeenergy}). Writing 
it in terms of $u$ and $v$, one has
\be
{\mathscr F}=\frac{\alpha}{2}\int_{-1}^1 dt\, (\dot{u})^2
e^v+\mu\int_{-1}^1dt\, e^{-v}, \label{freeenergy3}
\ee
We note that ${\mathscr F}$ is {\em exactly} of the form of a point
particle moving in a $0+1$-dimensional space-time. As such the full
action $\mathscr{S}+\mathscr{F}$ has the interpretation of a point
particle of mass $\mu$ interacting with two dimensional gravity. This
action is also a natural candidate for brane dynamics in
non-critical string theory \cite{Teschner:2003qk}. For our purposes, the condition
$\frac{d}{dt}u=\frac{d}{dn}v$ can be seen as a gauge choice of
parametrization of the worldline of the particle. Uniqueness of this
parametrization implies that the worldline of the particle is a simple
curve inside the domain. We note that the addition of ${\mathscr F}$
means that the function $v(w,\bar{w})$ defined above is no longer
harmonic, meaning the breakage of the single analytical chart
parametrization of the whole disk. Since the function is still
harmonic at $D_\pm$, one can define analytic maps in these domains
which thus will comprise the atlas of $D$ as alluded in the second Section. 

From the studies of Liouville field theory on a disk, the action
$\mathscr{S}+\mathscr{F}$ can be alternately interpreted as a
deformation of pure Liouville by means of vertex operators defined by
(\ref{freeenergy3}). These are exactly $E$ and $L$ defined in
(\ref{energy}) and (\ref{length}), albeit in a different
parametrization. It also follows that they satisfy the same OPE with
the stress-energy tensor as computed above and hence the critical
exponent should hold. The effect of the boundary is just to implement
a identification between vertex operators $V_\beta$ and $V_{Q-\beta}$
\cite{Fateev:2000ik}. The two terms
that comprise the potential in (\ref{stress}), implementing the
constraint that $z(w=\pm 1)=\pm 1$, are exactly of this
form. The effect of the boundary is then to render one of them
redundant, and thus the last ingredient of the action is the addition
of the term $\lambda \int d^2z\, e^{u+iv}$. This constraint can
nevertheless be dealt with by considering correlation functions of
fields inserted at $w=\pm 1$.

Given all the similarities explained above, one can thus posit that
the transition for large wire length $L\goesto L_0$ can be understood
in terms of the OPEs of vertex operators (see \cite{Thorn:2002am} for
a survey). In fact, since in this regime the domain is covered in
teardrop-like cells, one can conjecture that the system will undergo a
Kosterlitz-Thouless type of transition by coalescence of
``semi-vortices'' (or vortices dipoles) where
$\theta(z-i\epsilon)=\theta(z+i\epsilon)+\pi$, corresponding to the
point where the cell touches itself.  Another interesting prospect is
the interpretation of the action as gravity. It is believed that the
non-conformal coupling may help gravity fluctuations that are
important for inflation and dark energy \cite{Gutperle:2003xf}, thus we
hope to address this point in the simplified model proposed here in
future work.


{\it Acknowledgements ---} The author would like to thank
G. L. Vasconcelos, P. Lederer and especially M. A. F. Gomes for
discussions and suggestions. The author acknowledges partial support
from FACEPE and PROPESQ-UFPE.

\end{document}